\documentclass[useAMS,usenatbib,usegraphicx]{mn2e}

\title[A Model for the Offsets between X-ray and Radio Emission 
from Large Scale AGN Jets]
{A Model for the Offsets between X-ray and Radio Emission from 
  Large Scale AGN Jets}
\author[M. G. Kim and F. Takahara]
{M. G. Kim$^{1}$\thanks{E-mail: mgkim@vega.ess.sci.osaka-u.ac.jp}
  and F. Takahara$^{1}$ \\
  $^{1}$Department of Earth and Space Science,
  Graduate School of Science, Osaka University,
  Toyonaka, Osaka 560-0043, Japan}
\begin{document}

\date{}

\pagerange{\pageref{firstpage}--\pageref{lastpage}} \pubyear{2009}

\maketitle

\label{firstpage}

\begin{abstract}
  We investigate apparent internal structure of kiloparsec scale jets
  of AGNs arising from the energy dependent cooling of accelerated
  electrons and light travel time effect for relativistically moving
  sources. Using a simple cylindrical shell model, we find that the
  offsets between the peaks of X-ray, optical and radio brightness
  distributions observed for many cases are basically explained
  with this model. Assuming that electrons in the
  moving shell are instantaneously accelerated, X-rays are
  emitted for a shorter time scale and observed at the side nearer
  to the nucleus, while radio emission continues to the far side of
  the nucleus because of longer cooling time. The resultant offset
  turns out to be order of the jet diameter owing to the light travel
  time effect.
\end{abstract}


\section{Introduction}

It is well known that kiloparsec scale jets of radio loud AGNs
consist of many bright knots seen in X-ray and optical in addition
to the radio bands. For this decade the
{\it Chandra X-Ray Observatory} has detected large scale jets for
tens of objects with a sub-arcsecond resolution. Combining with radio
and optical observations with similar spatial resolution, offsets
between different observation bands have been noticed, e.g., for
M87 \citep{mar02, wil02}, Cen A \citep{kra02, har03}, 
PKS 1127-145 \citep{sie02}, 3C 66B \citep{har01} and
3C 31 \citep{har02}.
It is intriguing that almost always the centroids of X-ray knots are
located closer to the core than those of lower frequencies.
In M87 \citep{wil02, mar02}, the centroids of optical knots
are closer to the core than their radio counterparts
(for a recent review, see \citet{harr06}).
While the emission mechanism of radio through optical emission
is believed to be synchrotron radiation by relativistic electrons,
several mechanisms are proposed for X-ray emission.
X-rays from the knots of FR I galaxies such as
M87, Cen A, 3C 66B and 3C 31 are well interpreted with synchrotron
radiation by multi-TeV energy electrons, while for 
quasars such as PKS 1127-145 the mechanism has not 
been settled yet.
It may be inverse Compton scattering of CMB photons by
lower energy electrons with the aid of strong relativistic beaming
effect even on this large scale \citep{sie02, sie07}.

In this paper, assuming that X-ray emission is
synchrotron radiation by high energy electrons we discuss the
origin of the observed offsets and address the question why
in all cases, knots at higher frequencies are located closer to
the core than their counterparts at lower frequencies.
\citet{bai03} suggested that time lags due to synchrotron
losses can make these X-ray/optical and X-ray/radio offsets.
In the shock acceleration scenario, electrons in large scale jets 
can be
accelerated to energies high enough to emit synchrotron X-rays.
As they propagate downstream along the jet,
electrons cool as they emit at progressively lower frequencies
and result in radio/X-ray offsets at bright knots, e.g.,
the centroids of X-ray knots should be observed closer to the core.
In this case, however, we expect to see a time delay from
high to low frequencies if cooling effect is the sole cause of
the offsets. It is also noted that cooling time of radio emitting
electrons is too long to explain the observed offset.
In reality, X-ray/radio offsets
are seen at the same time for distant observers and
electron acceleration occurs for a finite time span and over a
finite spatial extent. Therefore, light travel time effects should
be taken into account, too.

This paper examines the possibility that X-ray/radio offsets
can be explained by a single
shock wave that accelerates electrons for a finite time span,
by considering the energy dependent cooling timescale of
relativistic electrons and light travel time effect of a
relativistically moving source with a finite spatial extent.
In \S 2 , we describe these effects and our model,
\S 3 and \S 4 present the results and discussion, respectively.
Finally, \S 5 summmarizes our conclusions.

\section{The model}

\subsection{Light travel time effect}

The light travel time effect is relevant when we consider
relativistically moving sources which have a finite spatial extent,
for example when a shock wave, or any kind of inhomogeneities,
travels along the jet. In such cases, emission from different
positions in the jet at the same coordinate time reaches the
observer at different time and will produce changes in the apparent
shape of the brightness distribution as viewed from the observer.
In other words, the apparent shape seen by the observer is different
from the shape of the emission region at the same coordinate time.
The difference is dependent on the viewing angle and the jet speed.

\begin{figure}
  \begin{center}
    \mbox{\includegraphics[bb=0 130 785 520,scale=.3,clip]
      {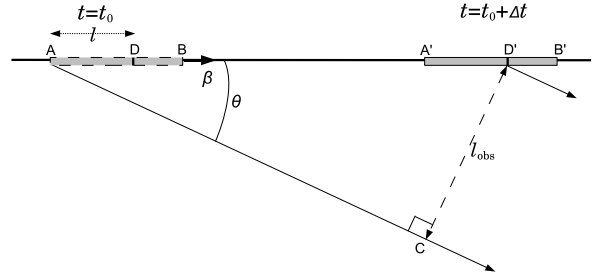}}
  \end{center}
  \caption{The light travel time effect 
    for the radiation emitted from an infinitesimally thin 
    rectilinear rod moving at a velocity of $\beta$ viewed at a
    viewing angle $\theta$.
    The emission region has a finite length from the point
    ${\rm A}$ to ${\rm B}$.
    The difference of arrival times at the observer
    between the radiation emitted from point A at time $t_0$ and
    that from point D at the time $t_0 + \Delta t$, represented as D',
    is different from $\Delta t$.}
    \label{fig:time_delay}
\end{figure}

First, we explain the simplest case shown in
Fig. \ref{fig:time_delay}, i.e., an infinitesimally thin rectilinear
rod moving at a velocity of $\beta$, normalized by the speed of light $c$, 
viewed at a viewing angle $\theta$.
The emission region has a finite length from the point ${\rm A}$
to ${\rm B}$. Let the point D be located at the distance $\ell$ from
the point A. The difference of arrival times at the observer
between the radiation emitted from point A at the time $t_0$ and
that from point D at the time $t_0 + \Delta t$ is expressed as
\begin{equation}
  \Delta t_{\rm obs} 
  = \Delta t - \left( \frac{\ell}{c} + \beta \Delta t \right)\cos\theta.
  \label{eq:timeobs}
\end{equation}
The condition of the same observer time $\Delta t_{\rm obs} = 0$
gives the time difference of
emission between the points A and D as
\begin{equation}
  \Delta t = \frac{\ell \cos \theta}{ c ( 1 - \beta \cos \theta )}.
  \label{eq:timediff}
\end{equation}
When the relativistic beaming effect is significant
($\cos\theta\approx \beta$), the time difference $\Delta t$ becomes
larger than $\ell/c$ about by a factor of $\Gamma^2=1/(1-\beta^2)$.
Moreover,
the apparent distance between points A and D, $\ell_{\rm obs}$, is
\begin{equation}
  \ell_{\rm obs}=(\ell+\beta c\Delta t)\sin\theta
  =\frac{\ell\sin\theta}{1-\beta\cos\theta}.
\end{equation}
This is also significantly larger than $\ell$
for relativistically beamed sources.
It is noted that this effect is also important for unbeamed sources
as far as the cooling time scale is shorter than $\Delta t$ 
in equation (\ref{eq:timediff}).
Interpretation of the jet structure should take into account this
light travel time effect. In reality the thickness of the jet
should be taken into account as is described in \S 2.3.

\subsection{Effects of radiative cooling of high energy electrons}

Another important aspect is the effect of the finite
cooling time of relativistic electrons.
It is widely accepted that electrons are accelerated by 
shocks in the jet.
In the framework of diffusive shock acceleration
\citep{bla87, bel78}, the time needed to accelerate electrons to a
Lorentz factor $\gamma$ in the jet comoving frame is
\citep{ino96, kus00}
\begin{equation}
  t_{\rm acc}( \gamma )
  = \frac{20 \lambda (\gamma) c}{ 3 u_{\rm s}^{2} }
  \sim \xi \left( \frac{\gamma}{10^7} \right)
  \left( \frac{B}{\rm 1G} \right)^{-1} {\rm sec},
  \label{eq:acc_timescale}
\end{equation}
where $u_{\rm s} \sim c$ is the shock speed,
$B$ is the magnetic field strength,
$\lambda (\gamma) = \xi \gamma m_{\rm e} c^2 / ( e B )$ is the mean
free path of electrons assumed to be proportional to the electron
Larmor radius with $\xi$ being a parameter
($\xi \gid 1$ and $\xi =1$ corresponds to the B\"ohm limit),
$m_{\rm e}$ is the electron mass, and $-e$ is the electron charge.
This equation implies that only several thousand seconds are needed
to accelerate electrons to the energy for which electrons emit
synchrotron X-ray radiation in 1 mG magnetic field.
The radiative cooling time $t_{\rm cool}(\gamma)$ of relativistic
electrons through synchrotron radiation is
\begin{equation}
  t_{\rm cool}(\gamma)
  = \frac{3 m_{\rm e} c}{4 \sigma_{\rm T} }U_{\rm B}^{-1}\gamma^{-1},
  \label{eq:cooling timescale}
\end{equation}
where $\sigma_{\rm T}$ is the Thomson cross section and
$U_B = B^2 / ( 8 \pi )$ is the energy density of the magnetic field.

In the jet comoving frame, the typical synchrotron emission frequency
of relativistic electrons, averaged over pitch angle, is
\begin{equation}
  \nu_{\rm s} = \frac{3}{4} \nu_B \gamma^2
  = \frac{3 e}{8 \pi m_e c} B \gamma^2
  \label{eq:relation_nu_gamma}
\end{equation}
where $\nu_{\rm B} = e B / (2 \pi m_{\rm e} c)$ is Larmor frequency.
In terms of $\nu_{\rm s}$, equation (\ref{eq:cooling timescale})
can be expressed as
\begin{eqnarray}
  t_{\rm cool}
  & \simeq & 2 \delta^{1/2} \left( \frac{B}{\rm 1mG} \right)^{-3/2}
  \left( \frac{\nu_{\rm obs}}{\rm 10^{17}Hz} \right)^{-1/2}
  {\rm yr}
  \label{eq:cooling_time}
\end{eqnarray}
where $\nu_{\rm obs} = \delta  \nu_{\rm s}$ is the typical synchrotron
frequency in the observer frame and
$\delta = [ \Gamma ( 1 - \beta \cos \theta ) ]^{-1}$
is the Doppler factor.
For example, the cooling time of high energy electrons emitting
$10^{17}$Hz photons
and that of low energy ones emitting $1$GHz photons,
differs by four orders of magnitude.
Thus, high energy electrons tend to be spatially confined
near the shock front, while low energy electrons tend to be
distributed in a more extended region.

\subsection{The model}

\begin{figure}
  \begin{center}
    \mbox{\includegraphics[bb=35 50 680 539,scale=.32,clip]
      {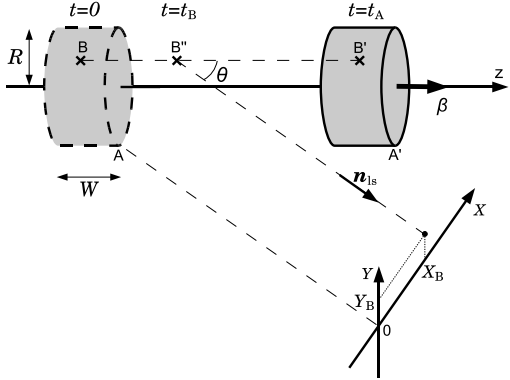}}
  \end{center}
  \caption{
    The emission from a shell with radius $R$ and width $W$
    moving at a constant velocity $\beta$ and
    the brightness distribution in the sky plane observed
    far from the jet with a viewing angle $\theta$.
    The shell begins to emit radiation at $t=0$ 
    and moves till $t=t_A$ along the $z$-axis.
    The point A represents the nearest position of the shell along the
    line of sight, while the point B represents an arbitrary position
    in the shell.
    The radiation emitted from the point B in the shell
    at the time $t=t_{\rm B}$ (the position designated by ${\rm B''}$)
    is observed at $(X_{\rm B}, Y_{\rm B})$ in the sky X-Y plane 
    which is set transverse to the line of sight.
    The origin $(X, Y) = (0, 0)$ is the position 
    where the radiation emitted from the position A at $t=0$ is observed.}
  \label{fig:shell}
\end{figure}

To investigate the above two effects, as the simplest model
we consider the emission from a shell with radius $R$ and width $W$
moving at a constant bulk Lorentz factor $\Gamma$ and
investigate the brightness distribution in the sky plane observed
far from the jet with a viewing angle $\theta$.
Hereafter, we adopt the cylindlical coordinate $(r, \phi, z)$ with
$z$-axis being the direction of the bulk motion as shown in
Fig. \ref{fig:shell}.
Let us consider the time delay between two points A and B in the shell,
the coordinates of which at time $t=0$ are given by
$(r_{\rm A}, \phi_{\rm A}, z_{\rm A})$ and
$(r_{\rm B}, \phi_{\rm B}, z_{\rm B})$, respectively.
The point A represents the nearest position of the shell along the
line of sight, while the point B represents an arbitrary position
in the shell.
We assume that at $t=0$ the shell begins to emit radiation
and that the emission continues till $t=t_{\rm cool}$.
In this case, the time delay between the radiation emitted from
the point B in the shell
at the time $t=t_{\rm B}$ (the position designated by ${\rm B''}$)
and the radiation emitted from the point
A in the shell at the time
$t=t_{\rm A}$ (the position designated by ${\rm A'}$) is
\begin{equation}
  \Delta t_{\rm obs}
  = (t_{\rm A} - t_{\rm B})
  + \mathbf{D} \cdot \mathbf{n_{ls}} / c,
  \label{eq:condition_t}
\end{equation}
where $\mathbf{D} \equiv \overrightarrow{\rm A'B''}$ is the vector
connecting the positions
${\rm A'}(r_{\rm A}, \phi_{\rm A}, c \beta t_{\rm A} + z_{\rm A})$
and
${\rm B''}(r_{\rm B}, \phi_{\rm B}, c \beta t_{\rm B} + z_{\rm B})$
and $\mathbf{n_{\rm ls}}$ is the unit vector toward the observer
along the line of sight.
For definition of the plane of sky,
we set the sky plane coordinates transverse to the line of sight 
as $(X, Y)$
and set the unit vectors as
$\mathbf{n_X} \equiv
(\mathbf{n_z} - (\mathbf{n_z} \cdot \mathbf{n_{ls}}) \mathbf{n_{ls}})
/ |\mathbf{n_z}-(\mathbf{n_z} \cdot \mathbf{n_{ls}}) \mathbf{n_{ls}}| $
and
$\mathbf{n_Y} \equiv
\mathbf{n_{ls}} \times \mathbf{n_X}
/ |\mathbf{n_{ls}} \times \mathbf{n_X} | $
, where $\mathbf{n_z}$ is the unit vector along the $z$-axis.
The origin $(X, Y) = (0, 0)$ on the X-Y plane is taken as 
the position where the radiation
emitted from the point $A$ at $t=0$ is observed.
The point ${\rm B''}$ is  observed at $(X_{\rm B}, Y_{\rm B})$,
where
\begin{eqnarray}
  X_{\rm B} &=& (\mathbf{D}+c\beta t_{\rm B}\mathbf{n_z}) \cdot \mathbf{n_X} 
  \label{eq:location_X} \\
  Y_{\rm B} &=& (\mathbf{D}+c\beta t_{\rm B}\mathbf{n_z}) \cdot \mathbf{n_Y}
  \label{eq:location_Y}
\end{eqnarray}

By calculating equations (\ref{eq:location_X}) and
(\ref{eq:location_Y})
under the  condition $\Delta t_{\rm obs} = 0$,
we determine the position and time of the point ${\rm B''}$
which is observed simultaneously with the point ${\rm A'}$.
Considering the coordinates of point B take a range of
$0 \lid r_{\rm B} \lid r_{\rm A}, 0 \lid \phi_{\rm B} \lid 2 \pi, 
0 \lid z_{\rm B} \lid z_{\rm A}$,
the brightness distribution on the plane of sky can be calculated.

In this paper, in order to demonstrate the effects of energy dependent
radiative cooling
we consider the simplest case where electrons have only two different
energies.
Lower energy electrons have longer cooling time $t_{\rm cool,low}$
and higher energy electrons shorter one $t_{\rm cool,high}$.
Moreover, we treat $t_{\rm cool,low}$ and $t_{\rm cool,high}$ as
phenomenological parameters and
simply assume that electrons radiate at a constant rate for
the respective time spans.

As for the particle acceleration,
we ignore the acceleration timescale
and assume that electrons of all energy  begin to emit synchrotron
radiation at the same time $t=0$,
because the acceleration timescale is much shorter than the cooling
timescales.

Of course, this model is just a toy model to demonstrate effects
of light travel time and energy dependent radiative cooling.
To improve on more realistic cases is not difficult once these
effects are understood.

\subsection{Analysis in the $Y={\rm const.}$ plane}

\begin{figure}
  \begin{center}
    \mbox{\includegraphics[bb=0 0 785 550,scale=.3,clip]
      {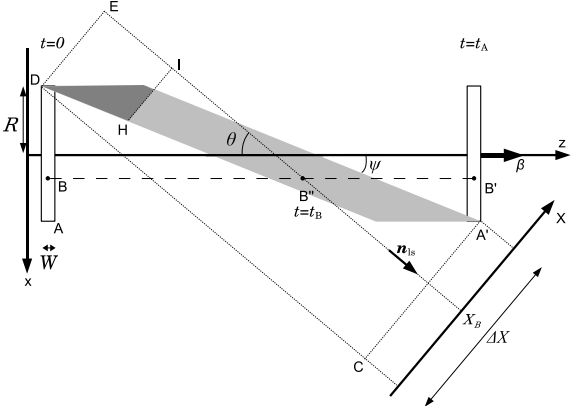}}
  \end{center}
  \caption{
    The observable region of a moving shell in the two-dimensional plane
    containing the vector $\mathbf{n_{\rm ls}}$ and $\mathbf{n_{\rm z}}$
    We set the $x$-axis transverse to the $z$-axis.
    The radiating shell moves along the $z$-axis from left to right
    and both light and dark grey-colored regions represent the
    region observed simultaneously.
    The region is segmented by a range of source time such that for more
    distant part along the line of sight the emission time is earlier.
    The dark grey-colored region represents
    the region of high frequency radiation, while
    the light grey-colored region represents
    the region of low frequency radiation.}
  \label{fig:2d_shell}
\end{figure}

Owing to the energy dependent cooling, an offset
between the brightness distributions of high and low frequency
emission arises when the light travel time and cooling effects
are taken into account.
Here, we discuss the typical magnitude expected in the 
two-dimensional plane
containing the vector $\mathbf{n_{\rm ls}}$ and $\mathbf{n_{\rm z}}$,
as seen in Fig. \ref{fig:2d_shell}. In the sky plane, this plane is
projected onto the line of $Y=$ const.
We set the $x$-axis transverse to the $z$-axis
to describe the position of the shell material and generarize the
consideration made in section 2.1 to the shell with a finite thickness.
Since $\mathbf{n_{\rm ls}}=(\sin\theta,\cos\theta)$ and
$\mathbf{D}
=(x_{\rm B}-x_{\rm A},z_{\rm B}-z_{\rm A}+c\beta(t_{\rm B}-t_{\rm A}))$,
using equation (\ref{eq:condition_t}) the condition of
simultaneous observation $\Delta t_{\rm obs}=0$ gives
the difference of the emission time for the points A and B, 
$\Delta t = t_{\rm A} - t_{\rm B}$, as 
\begin{equation}
  \Delta t =
  \frac{\Delta x \sin \theta + \Delta z \cos \theta}
  {c ( 1 - \beta \cos \theta )}
  \label{eq:2d_condi}
\end{equation}
where 
$\Delta x = x_{\rm A} - x_{\rm B}$ and
$\Delta z = z_{\rm A} - z_{\rm B}$.
So the maximum value of $\Delta t$ is expressed as
\begin{equation}
  {\rm Max}(\Delta t)
  = \frac{2R\sin\theta + W\cos\theta}{c(1 - \beta \cos \theta )}
  \label{eq:2d_condi_max}.
\end{equation}

The position and time of the point
${\rm B''}(x_{\rm B''}, z_{\rm B''})$
observed simultaneously with point ${\rm A'}$ at $t=t_{\rm A}$
are given respectively by using the relation
$z_{\rm B''} = z_{\rm B} + c \beta t_{\rm B}$ and $x_{\rm B''}
= x_{\rm B}$
as
\begin{equation}
  z_{\rm B''}
  = \frac{(x_{\rm B}-x_{\rm A}) \beta \sin \theta + z_{\rm B}
    - z_{\rm A} \beta \cos \theta}
  {1 - \beta \cos \theta  }
  + c \beta t_{\rm A}
  \label{eq:2d_condi_02}
\end{equation}
and
\begin{equation}
  t_{\rm B}
  = \frac{(x_{\rm B} - x_{\rm A})\sin \theta }{c}
  + \frac{(z_{\rm B''} - z_{\rm A})\cos \theta }{c}
  + t_{\rm A}(1-\beta \cos \theta) .
  \label{eq:2d_condi_03}
\end{equation}

The coordinates of $x_{\rm B}$ and $z_{\rm B}$ take a
range of $-R \lid x_{\rm B} \lid R =x_{\rm A}$
and $0 \lid z_{\rm B} \lid W=z_{\rm A}$,
so the region observed simultaneously
can be drawn
by equation (\ref{eq:2d_condi_02})
as the grey-colored region in Fig. \ref{fig:2d_shell}
for a given value of $t_{\rm A}$.
In Fig. \ref{fig:2d_shell},
the radiating shell moves along the $z$-axis (from left to right)
and both light and dark grey-colored regions represent the
regions observed simultaneously.
The region is segmented by a range of source time $t_{\rm B}$ 
such that for more
distant part along the line of sight the source time is earlier.
The dark grey-colored region represents
the region of high frequency radiation, while
the light grey-colored region represents
the region of low frequency radiation.
It should be noted that in Fig. \ref{fig:2d_shell}
we assume that the particle acceleration occurs instantaneously
$t=0$ and synchrotron radiation begins at the time $t = 0$,
and we set the time $t_{\rm A} = {\rm Max}(\Delta t) $ and
the cooling timescales satisfy
$t_{\rm cool,high} < t_{\rm A} < t_{\rm cool,low}$.

The lines DE, HI and C${\rm A'}$ in Fig. \ref{fig:2d_shell} 
represent those of equal source time
$t=0, \ t_{\rm cool,high} \ {\rm and} \ t_{\rm A}$,
respectively.
So both the light and dark grey regions are observable at the 
low frequency
while at the high frequency only the dark grey region is observed.

Because we assume that the emissivity in the shell is independent
of time as far as $t\le t_{\rm cool}$,
the brightness on the plane of sky is simply proportional
to the length of the emitting region along the line of sight.
Then we can calculate the brightness distribution along the $X$-axis
and see the offset between the brightness distributions between 
the high and low
frequencies.

The inclination of the hatched region is given by 
\begin{equation}
  \tan\psi = \frac{1 -\beta \cos \theta}{\beta \sin \theta} ,
\end{equation}
so that $\theta > \psi$ when $\sin\theta > 1/\Gamma$
and $\theta < \psi$ when $\sin\theta < 1/\Gamma$.
Fig. \ref{fig:2d_shell} depicts the former case and 
the apparent size of the shell seen in the low frequency is given by 
\begin{equation}
  \Delta X
  = \frac{2R \mid \beta - \cos \theta \mid +
    W \sin \theta}{1-\beta \cos \theta}.
  \label{eq:delta_x}
\end{equation}
For the latter case, it is also given by equation (\ref{eq:delta_x}).
We notice the apparent size gives the upper limit on the offset. 

\section{Results}\label{sec:result}

Here we present the brightness distribution on the plane of sky
at two frequencies corresponding to two energies of electrons.
Especially, the cases that the viewing angle is around $1/\Gamma$ are
investigated.

\begin{figure}
  \begin{center}
    \mbox{\includegraphics[bb=80 50 320 220,scale=0.9,clip]
      {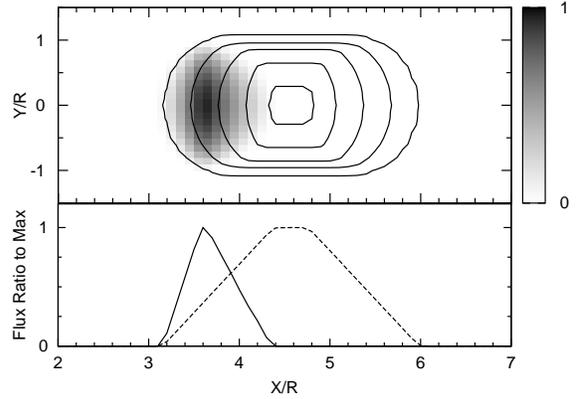}}
  \end{center}
  \caption{
    Top panel: The brightness contour of the radiation
    emitted from the shell which is observed at the observer time 
    $t_{\rm obs} \simeq 0.75R/c$.
    The $X$ and $Y$ axes are the same as in Fig. \ref{fig:shell}.
    We set $t_{\rm A}=30 R/c$, 
    $\sin\theta = 2/\Gamma$,
    $t_{\rm cool,low} = 60 R/c$, 
    $t_{\rm cool,high} = 0.2 t_{\rm cool,low}$,
    $W/R = 0.2$ 
    and $\Gamma = 10$.
    The solid line contours show the brightness distribution
    from the low energy electrons,
    while the gray-colored contour shows that from the high energy ones.
    The line contours represent from outside to inside 2\%,
    25\%, 50\%, 75\%, and 95\% of the peak value.
    The density of the gray contour shows the ratio of the brightness to
    its peak value.
    Bottom: The brightness distribution along the $Y=0$ axis in the 
    top panel.
    The solid line shows the high energy flux 
    and broken line does the low energy one.
    The magnitude of the offset between the peaks of two 
    distributions is $D_{\rm off} \sim 1.0 R$.}
  \label{fig:result_1}
\end{figure}

Fig. \ref{fig:result_1} shows the brightness contour of the 
synchrotron radiation
emitted from the shell which is observed at the observer time
$t_{\rm obs} \simeq 0.75 R/c$ 
where we set the source time $t_{\rm A}=30 R/c$,
viewing angle $\sin\theta = 2/\Gamma$,
cooling times $t_{\rm cool,low} = 60 R/c$, 
$t_{\rm cool,high} = 0.2 t_{\rm cool,low}=12R/c$,
the ratio $W/R = 0.2$
and bulk Lorentz factor $\Gamma = 10$.
By equation (\ref{eq:timeobs}) we set the observer time 
$t_{\rm obs}= t_{\rm A} (1-\beta \cos\theta)$
and $t_{\rm obs}=0$ represent the time when the emission from the 
point A at $t=0$ is observed.
The solid line contours show the brightness distribution
from the low energy electrons,
while the gray-colored contour shows that from the high energy ones.
The line contours represent 2\%,
25\%, 50\%, 75\%, and 95\% of the peak value  from outside to inside .
The density of the gray contour shows the ratio of the brightness to
its peak value.
Obviously, the offset between the centroid of the line contours and
that of the gray-colored region is seen.
The magnitude of the offset turns out to be $D_{\rm off} \sim 1.0 R$
for this case.

\begin{figure}
  \begin{center}
    \mbox{\includegraphics[bb=80 50 320 220,scale=0.9,clip]
      {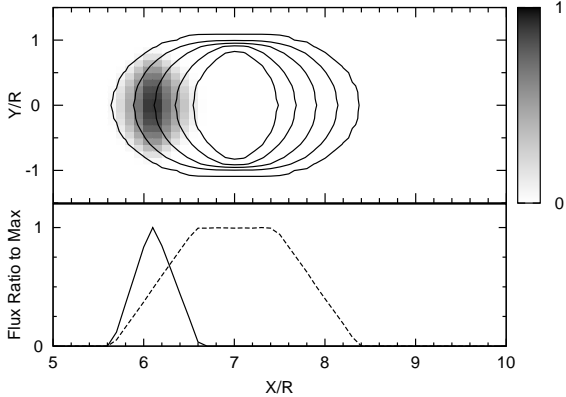}}
  \end{center}
  \caption{
    Top panel: The brightness contour 
    which is observed at the observer time 
    $t_{\rm obs} \simeq 4.3R/c$
    We set $t_{\rm A}=21 R/c$, 
    $\sin\theta = 4/\Gamma$,
    $t_{\rm cool,low} = 60 R/c,
    t_{\rm cool,high} = 0.2 t_{\rm cool,low}$,
    $W/R = 0.2$
    and $\Gamma = 10$.
    Bottom: The brightness distribution along the $Y=0$ axis in the top panel.
    The magnitude of the offset between the peaks of two 
    distributions is $D_{\rm off} \sim 0.9 R$.}
  \label{fig:result_3}
\end{figure}

The offset is not due to the beaming effect \citep{ryb79} 
but the relativistic motion of jet.
That is showed in Fig. \ref{fig:result_3} where the velocity of jet is 
the same as that in Fig. \ref{fig:result_1} but the viewing angle 
is much larger than $1/\Gamma$. 
Fig. \ref{fig:result_3}  shows the brightness contours
at the observer time $t_{\rm obs} \simeq 4.3R/c$
for the case of $t_{\rm A}=21 R/c$, $\sin\theta = 4/\Gamma$,
$t_{\rm cool,low} = 60 R/c$, $t_{\rm cool,high} 
= 0.2 t_{\rm cool,low}=12R/c$,
$\Gamma = 10$ and $W/R =0.2$.
In this case, the offset turns out to be  $D_{\rm off} \sim 0.9 R$
and almost the same as that in Fig. \ref{fig:result_1}.

\begin{figure}
  \begin{center}
    \mbox{\includegraphics[bb=80 50 320 220,scale=0.9,clip]
      {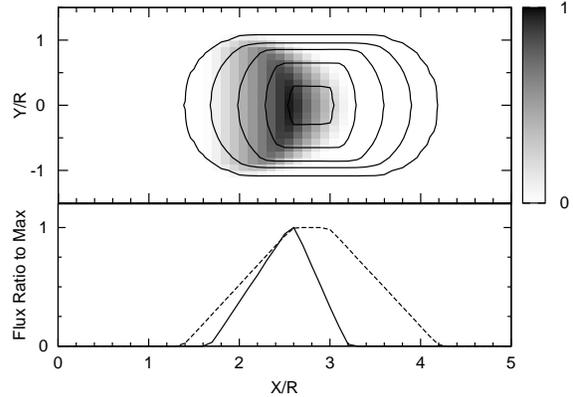}}
  \end{center}
  \caption{
    Top panel: The brightness contour 
    which is observed at the observer time 
    $t_{\rm obs} \simeq 0.3R/c$
    We set $t_{\rm A}=48 R/c$, 
    $\sin\theta = 1/(2\Gamma)$,
    $t_{\rm cool,low} = 60 R/c,
    t_{\rm cool,high} = 0.2 t_{\rm cool,low}$,
    $W/R = 0.2$
    and $\Gamma = 10$.
    Bottom: The brightness distribution along the $Y=0$ axis in the top panel.
    The peak of the high energy radiation is located inside of the region where 
    the brightness of the low energy radiation has a flat top peak,
    but has an offset $D_{\rm off} \sim 0.2 R$.}
  \label{fig:result_2}
\end{figure}

\begin{figure}
  \begin{center}
    \mbox{\includegraphics[scale=.19]
      {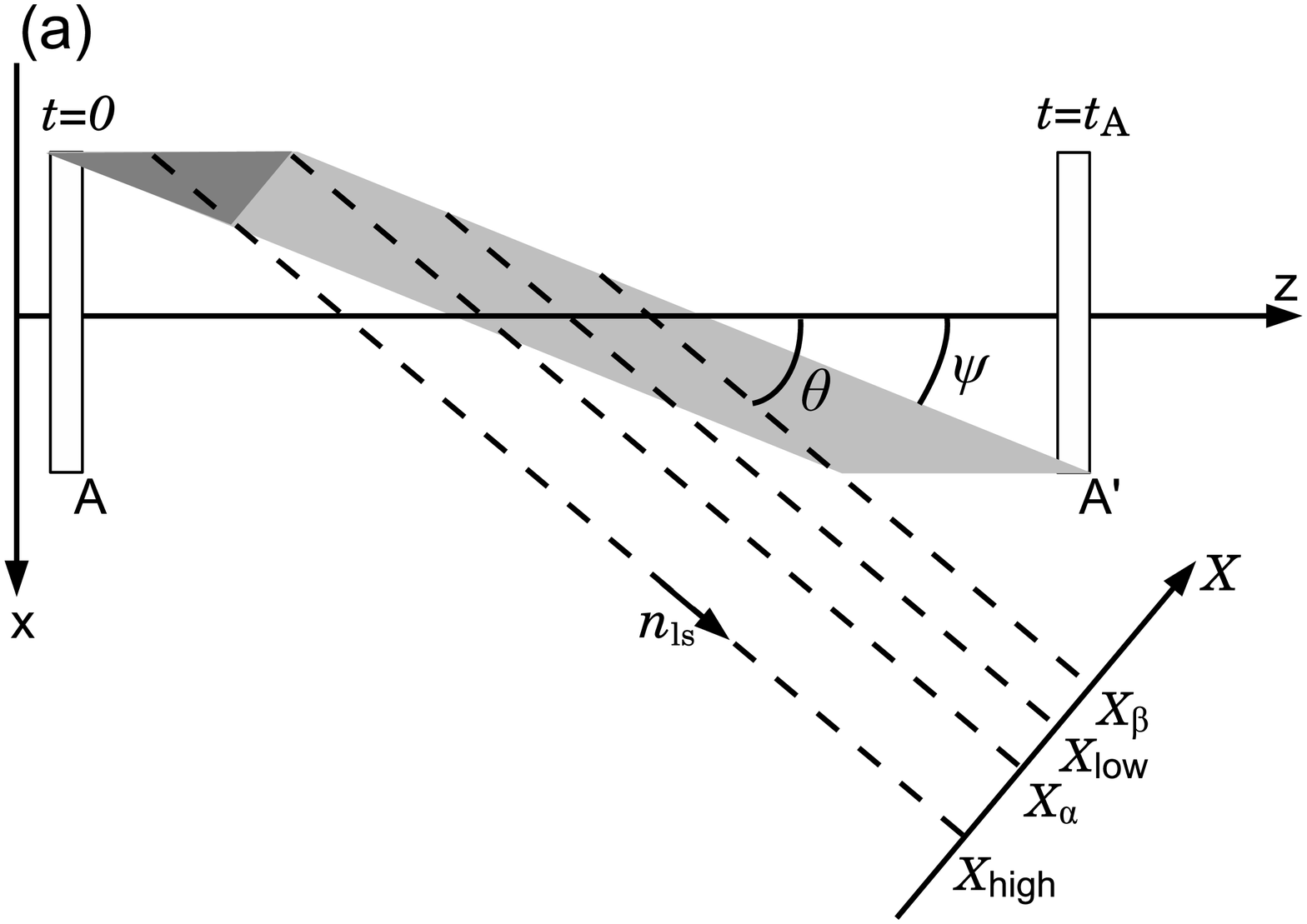}}
    \mbox{\raisebox{10mm}{\includegraphics[bb=50 0 700 550,scale=.13,clip]
      {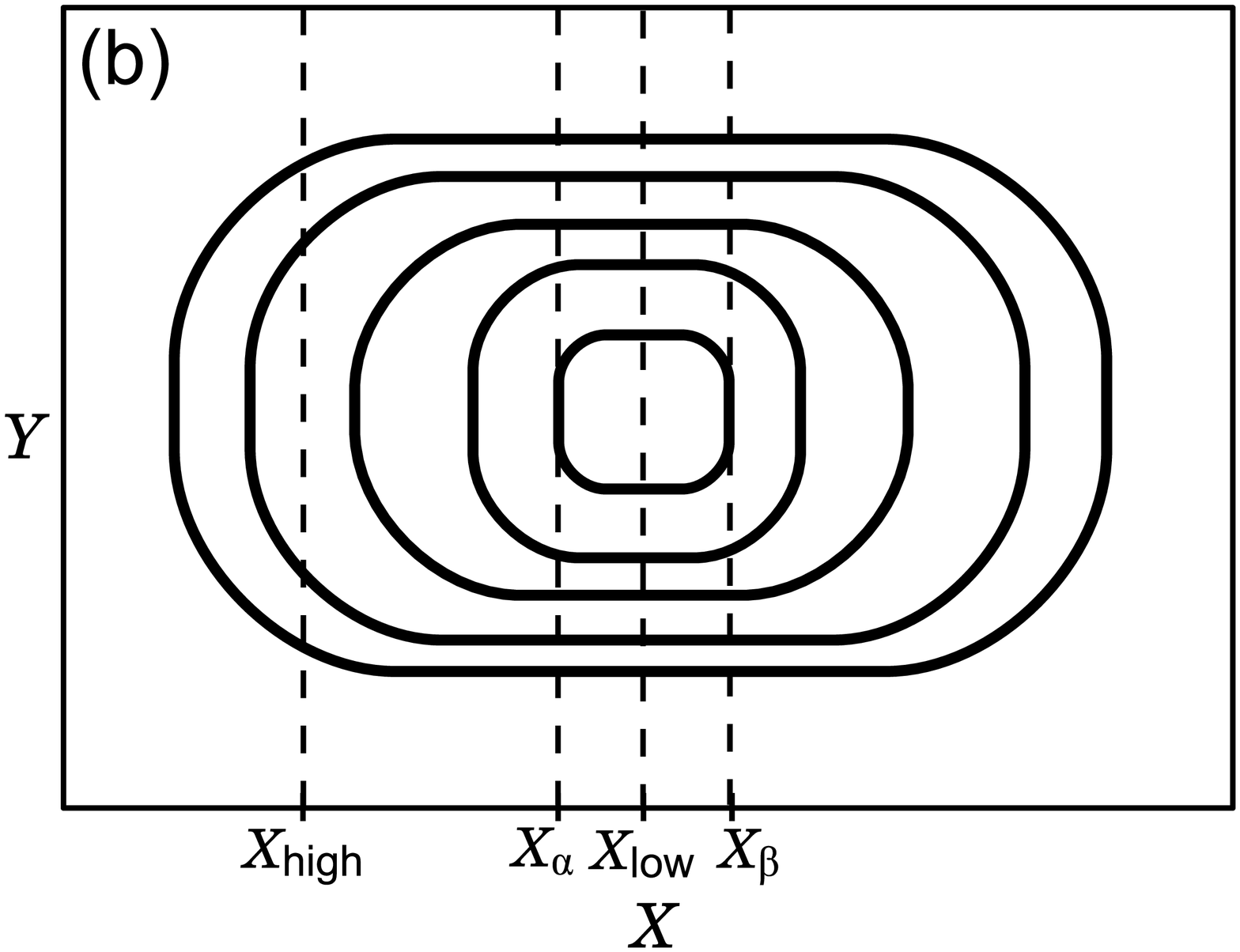}}}
  \end{center}
  \begin{center}
    \mbox{\includegraphics[bb=0 120 785 539,scale=.19,clip]
      {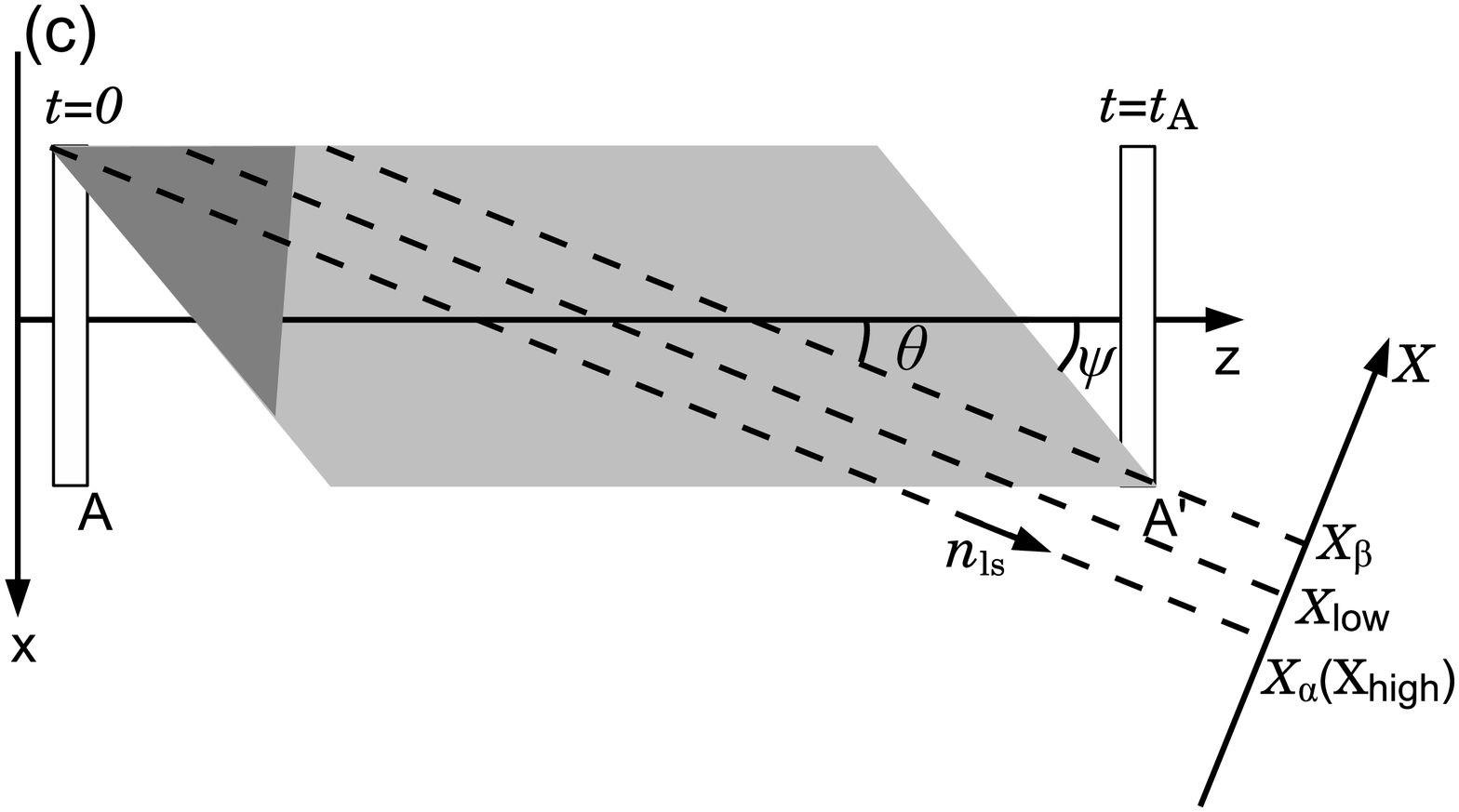}}
    \mbox{\raisebox{2mm}{\includegraphics[bb=50 0 700 550,scale=.13,clip]
      {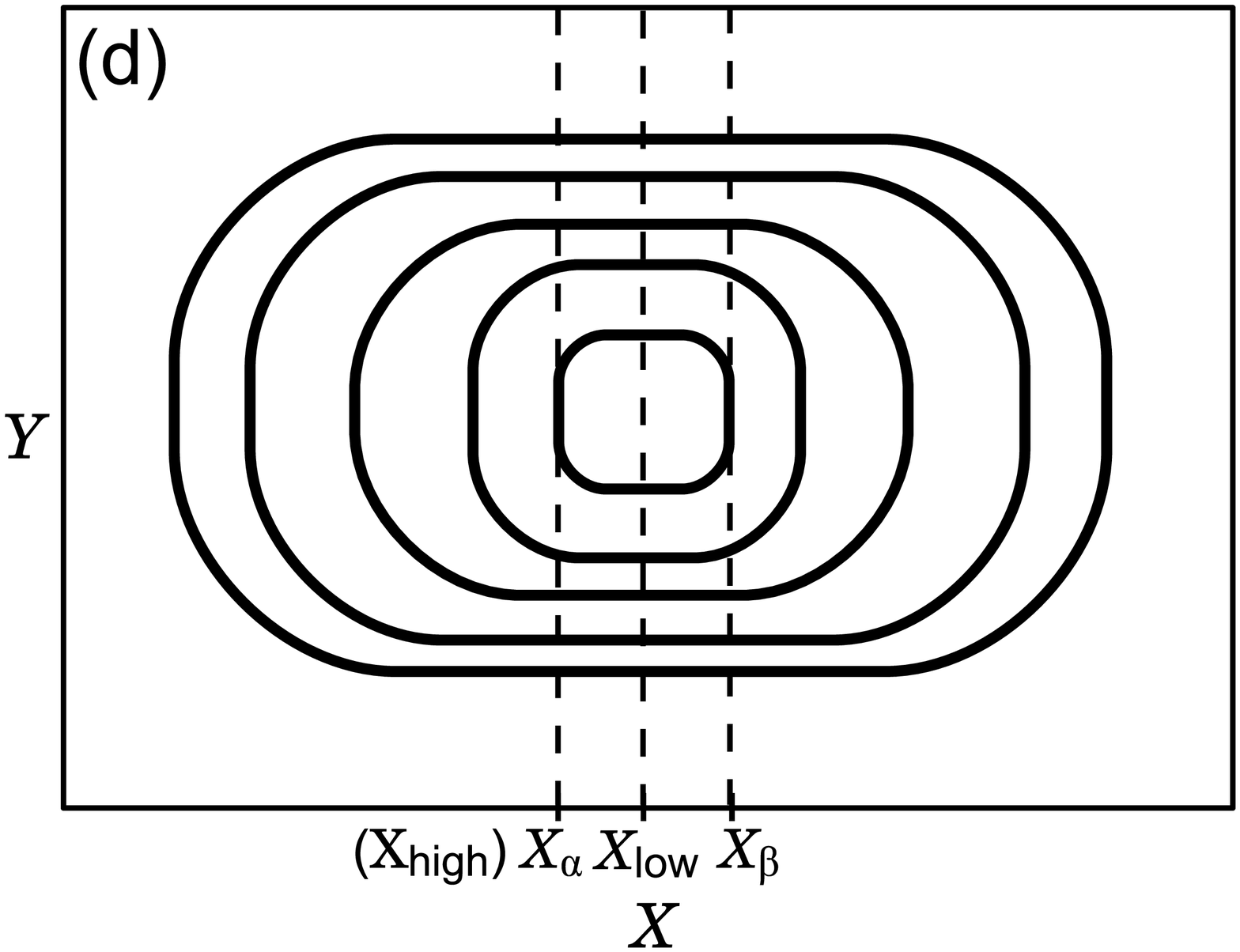}}}
  \end{center}
  \caption{
    The diffrence between the cases of $\sin\theta>1/\Gamma$ and 
    $\sin\theta<1/\Gamma$. The upper panels are for $\sin\theta > 1/\Gamma$, 
    while the lower panels are for $\sin\theta < 1/\Gamma$. 
    The left panels show the moving shell in the two-dimensional 
    plane and both light and dark grey-colored regions represent the
    region observed simultaneously as in Fig. \ref{fig:2d_shell}
    and the segmentation is also the same as in Fig. \ref{fig:2d_shell}.
    The right panels show the brightness contours of the low 
    energy radiation. 
    In both panels, the centroid of the high energy radiation is 
    marked by $X=X_{\rm high}$
   while  that of the low energy one is by $X=X_{\rm low}$,
   The brightness of the low energy radiation shows a flat top
    distribution  between $X=X_\alpha$ and $X_\beta$.}
  \label{fig:theta_var}
\end{figure}

The magnitude of the offset is dependent on the assumed parameters.
Fig. \ref{fig:result_2}  shows the brightness contours
at the observer time $t_{\rm obs} \simeq 0.3R/c$
for the case of $t_{\rm A}=48 R/c$, $\sin\theta = 1/2\Gamma$,
$t_{\rm cool,low} = 60 R/c$, $t_{\rm cool,high} 
= 0.2 t_{\rm cool,low}=12R/c$,
$\Gamma = 10$ and $W/R =0.2$.
In this case, the offset turns out to be  $D_{\rm off} \sim 0.2 R$.
The difference is largely due to the difference in viewing angle. 
As has been noted in the previous section, 
the case $\sin\theta=2/\Gamma$ the shell is seen from the tail side
while for $\sin \theta=1/(2\Gamma)$ it is seen from the front side.  
As is seen in Figs. \ref{fig:result_1} and \ref{fig:result_2}
larger offset is expected when  $sin\theta>1/\Gamma$, 
while the offset is small when  $sin\theta<1/\Gamma$.
Fig. \ref{fig:theta_var} describes a closer inspection 
of the situation. As is seen in the former case, the brightness peak
from the high energy electrons (the position designated by $X_{\rm high}$
is located outside the flat top peak of the brightness contour 
from the low energy ones (the segment desinated by $X_{\rm \alpha} X_{\rm \beta}$),
while for the latter case the position $X_{\rm high}$ is inside 
the segment $X_{\rm \alpha} X_{\rm \beta}$.
For the both cases, it is noticed the offset between the position 
$X_{\rm high}$ and $X_{\rm low}$, which represent the centroid 
of the brightness distribution from the low energy ones, exist.

\begin{figure}
  \begin{center}
    \mbox{\includegraphics[bb=75 118 255 152,scale=1.3,clip]
      {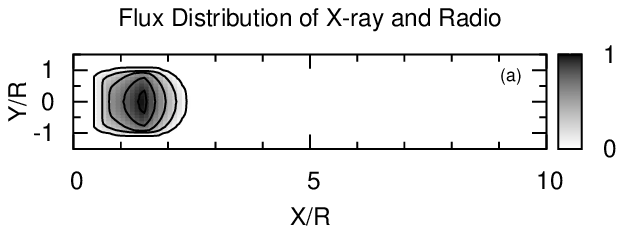}}
  \end{center}
  \begin{center}
    \mbox{\includegraphics[bb=75 118 255 152,scale=1.3,clip]
      {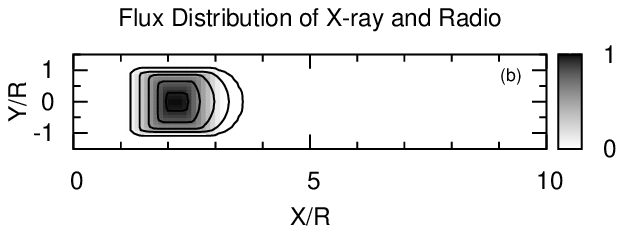}}
  \end{center}
  \begin{center}
    \mbox{\includegraphics[bb=75 118 255 152,scale=1.3,clip]
      {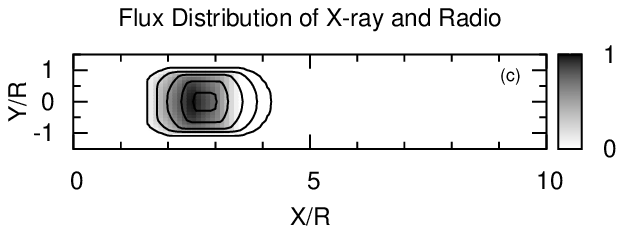}}
  \end{center}
  \begin{center}
    \mbox{\includegraphics[bb=75 118 255 152,scale=1.3,clip]
      {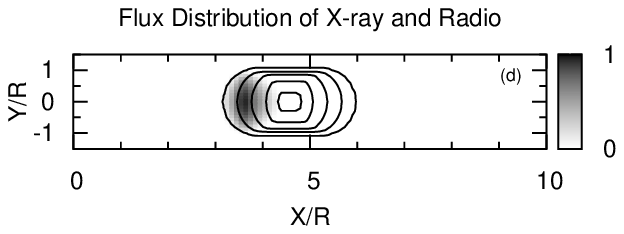}}
  \end{center}
  \begin{center}
    \mbox{\includegraphics[bb=75 98 255 152,scale=1.3,clip]
      {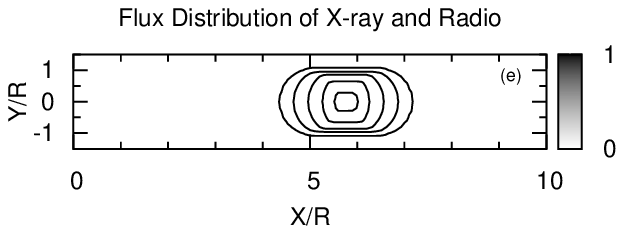}}
  \end{center}
  \caption{
    Time evolution of the brightness distribution on the plane of sky
    for the case shown in Fig. \ref{fig:result_1}.
    The profiles of line and greyscale contours are depicted in the 
    same way as in Fig. \ref{fig:result_1},
    and all the brightness distributions are normalized by the peak value
    for each epoch.
    The epoch for panel $(a) \sim (e)$ corresponds to 
    $t_{\rm obs}/(R/c) = 0.3, \ 0.45, \ 0.53, \ 0.75 \ {\rm and} \ 0.9$, 
    respectively.}
    \label{fig:time_var}
\end{figure}

The brightness profile is also time-dependent.
In Fig. \ref{fig:time_var},
the time variation of the brightness distribution on the plane of sky is
shown for the case shown in Fig. \ref{fig:result_1}.
The profiles of line contours and greyscale contours are shown in the 
same way as in Fig. \ref{fig:result_1},
and all the brightness distributions are normalized by the peak values
for each epoch.
The epoch for panels $(a) \sim (e)$ is
$t_{\rm obs}/(R/c) = 0.3, \ 0.45, \ 0.53, \ 0.75 \ {\rm and} \ 0.9$
corresponding to 
$t_{\rm A}/(R/c) = 12, \ 18, \ 21, \ 30 \ {\rm and} \ 36$, 
respectively.
As is seen in panels $(a)$ and $(b)$, 
the offset is not significant for early stage 
before higher energy electrons cool.
After high energy electrons have cooled,
the offset arises and becomes larger with time
as seen in panels $(c) \sim (e)$.
As a result, while the high energy radiation is observed for 
a time span of $0.90 R/c$, 
offset is observed from $0.53 R/c$ to $0.9 R/c$.
Thus, the offset is observed for about a half of the observing time 
of the high energy flux.
Moreover, the area covering the low frequency emission keeps increasing,
but that of the high frequency emission first increases and reaches
a maximum in panel $(b)$ and turns to decrease afterward.
This time variation may be one of the reasons why
the extent of the offsets is different in different knots,
including the cases of no offset.
After some time the shell emits only emission from lower energy electrons
and seen only in the radio band.

\section{Discussion} \label{sec:discussion}

\subsection{Statistics of the offsets}

We have shown that the offset between X-rays and radio emission 
naturally arises when we take account of the energy dependent 
radiative cooling of relativistic elctrons and the light travel time 
effect for moving sources. We estimated that roughly half of the 
X-ray knots should show an offset from the radio knots. 

In the above we adopted an artificially large ratio of the 
cooling time of X-ray emitting 
electrons to that of radio emitting electrons as 0.2 compared 
with the realistic value of around $10^{-4}$. 
This does not affect the above conclusion but 
affect the fraction in which X-ray knots are observed among 
radio knots. In the above we have assumed that particle acceleration 
is instantaneous but in reality particle acceleration occurs 
for a finite time span. The observability of X-ray knots 
is determined by the ratio of this time span to the cooling time 
of radio emitting electrons. In addition, adiabatic expansion 
of radio knots may play a role in determining the life time of 
radio knots.  

The ratio of the cooling time of radio emitting to 
the maximum time difference of emission time in the source 
${\rm Max}(\Delta t)$ is given by 
\begin{eqnarray}
  \frac{t_{\rm cool.low}}{{\rm Max}(\Delta t)}
  &=& \left(\frac{R}{6\times10^3 {\rm pc}}\right)^{-1}
  \left( \frac{B}{\rm 1mG} \right)^{-3/2}
  \left( \frac{\nu_{\rm obs}}{\rm 1GHz} \right)^{-1/2}\nonumber \\
  &&\times
  \left( \frac{1-\beta \cos \theta}{2\sin \theta + (W/R) \cos \theta} 
  \right) \delta^{1/2} .
  \label{eq:condition_R}
\end{eqnarray}
The order of magnitude of this value is of order unity, 
which is compatible with our model.  

\subsection{Implication for the indivisual offsets in M87 and Cen A}

Now we discuss on the application of our model to the offsets 
seen in typical objects.
First, we discuss on FRI radio galaxies such as M 87 and Cen A. 
Although for FRI radio galaxies viewing angle is relatively large and 
relativistic beaming effects are mild, 
the offset between X-ray and radio emission is expected as far as 
the knots are moving relativistically or even at mildly relativistic 
velocities. And its magnitude is a significant fraction of $\Delta X$.
As for M87 located at a distance of 16 Mpc and $1'' = 78 {\rm pc}$,
the observed radio/X-ray offsets are found as $0''.5$ for knot D and
$0''.7$ for knot F \citep{wil02},
while $\Delta X$ in our model is estimated as $2''.0$
for $\theta = 20^{\circ}$, $\Gamma = 10$ \citep{bir99},
$R = 58.5 {\rm pc}$ corresponding to $0''.75$ and $W/R = 0.2$. 
For $B =10^{-4}G$ \citep{wil02} and 
$R$ satisfies equation (\ref{eq:condition_R}) which means 
high energy electrons cool sufficiently fast.
It is interesting to note that the X-ray knots D and F are not 
separated from the radio knots but a part of the radio knots, 
which is fully consistent with our results shown in Figs. 
\ref{fig:result_1} and \ref{fig:result_2}.
Thus, our model is compatible with observations of M87.

In Cen A located at a distance of $3.4 {\rm Mpc}$ and 
$1'' = 17 {\rm pc}$,
the radio/X-ray offsets increase from $1''.3$ for knots A2A/AX2
to $3''$ for knots A3B/AX3A and $8''$ for B1A/BX2
\citep{har03}.
This feature is probably due to the increase of the jet radius $R$
with the distance from the nucleus
and we set $R = 2'', 3'', 14''$ \citep{kra02}
for respective knots corresponding to 
$R=34 {\rm pc},\ 51 {\rm pc} \ {\rm and} \ 238 {\rm pc}$.
These radii satisfies the cooling condition 
of equation (\ref{eq:condition_R}).
It should be noticed the viewing angle and velocity of the jet 
in Cen A are not well determined only with the constraints of 
$50^\circ < \theta < 80^\circ$ and $\beta > 0.45$ \citep{tin98, jon96}.
Thus, we examine  two sets of parameters.
First we set a large vieing angle and relativistic velocity such as  
$\theta = 80^\circ$,
$\Gamma = 10$,
$B = 10^{-4}G$ and 
$W/R =0.2$. For this set we obtain
$\Delta X$ of $4''.4, \  6''.7, \ 31''$ for respective knots.
(The differences of $\Delta X$ come from those of $R$.)
Second choise is a moderate viewing angle and moderate velocity such as 
$\theta = 50^\circ$,
$\Gamma = 1.1$,
$B = 10^{-4}G$ and 
$W/R =0.2$. For this case we obtain 
$\Delta X$ of $1''.0, \ 1''.6, \ 7''.2$ for respective knots and 
these are smaller than the observations.
Thus, the first case of 
larger viewing angle and faster velocity of jet are prefered.

Finally, we discuss on offsets of quasar jet in PKS 1127-145 
which located at $z=1.187$ and $1''$ corresponds to $11.5kpc$ \citep{sta98}.
The largescale radio/X-ray offsets are found as $2''.1$ for knot A,
$1''.4$ for knot B and $0''.8$ for knot C \citep{sie02}.
The jet radius $R$ is around $2''$ for every konts \citep{sie02} 
and satisfies equation (\ref{eq:condition_R}) for $B=1mG$.
However, the viewing angle and velocity of the jet are not well known yet 
\citep{sie02, sie07} so we examine parameters which meet the condition 
$\Delta_X > 2''$ by equation (\ref{eq:delta_x}).
Consequently we find when $\Gamma > 5$ or $\sin\theta <0.7/\Gamma$ or 
$\sin\theta>1.3/\Gamma$ the magnitudes of the offsets are comptible with 
our model. 
However, X-ray emission mechanism in these sources 
is still an open problem because the X-ray spectrum is 
definitely flatter than the extrapolation of radio to optical 
spectrum \citep{sie02}. 
In spite of this complexities, the offset may suggest 
that X-rays are from high energy electrons with short cooling time.

\section{Summary}

We investigated apparent internal structure of kiloparsec scale jets
of AGNs arising from the energy dependent cooling of accelerated
electrons and light travel time effect for relativistically moving
sources. 
Using a simple cylindrical shell model, we find that the
offsets between the peaks of X-ray, optical and radio brightness
distributions observed for many cases are basically explained. 
Assuming that electrons in the
moving shell are instantaneously accelerated, X-rays are
emitted for a shorter time scale and observed at the side nearer
to the nucleus, while radio emission continues to the far side of
the nucleus because of longer cooling time.
The resultant offset in Fig. \ref{fig:result_1}
turns out to be order of the jet diameter owing to the light travel
time effect.

The magnitude of the offset is dependent on the viewing angle $\theta$
and differs by one order of magnitude 
when $\sin\theta$ varies from $2/\Gamma$ to $1/(2\Gamma)$.
We showed that  the apparence of the offset 
alters at $\sin\theta = 1/\Gamma$ as shown in Fig. \ref{fig:theta_var}.
When $\sin\theta > 1/\Gamma$ we see the shell from the tail side,
while when $\sin\theta < 1/\Gamma$ we wee it from the front side.

We also investigated the time evolution of the brightness profile.
The offset is not significant for early stage 
before higher
energy electrons cool.
After the stage when high energy electrons have cooled,
the offset arises and becomes larger with time.
As a result, the offset is observed for about a half of the observing time 
of the X-ray knots.
This time variation may be one of the reasons why
the extent of the offsets is different in different knots,
including the cases of no offset.

Finally we argued that the observed  offsets in FR I radio galaxies M87 
and Cen A and other kpc jets of quasar PKS 1127-145 are well interpreted 
with our model.

\label{lastpage}


\begin{thebibliography}{99}
\bibitem[\protect\citeauthoryear{Bai \& Lee}{2003}]{bai03}
  Bai J. M., Lee M. G., 2003, ApJ, 585, 113
\bibitem[\protect\citeauthoryear{Bell}{1978}]{bel78}
  Bell A. R., 1978, MNRAS, 182, 147
\bibitem[\protect\citeauthoryear{Biretta, Sparks \& Macchetto}{1999}]{bir99}
  Biretta J. A., Sparks W. B., Macchetto F., 1999, ApJ, 520, 621
\bibitem[\protect\citeauthoryear{Blandford \& Eichler}{1987}]{bla87}
  Blandford R. D., Eichler D., 1987, Phys.Rep., 154, 1
\bibitem[\protect\citeauthoryear{Ghisellini et al.}{1998}]{ghi98}
  Ghisellini G., Celotti A., Fossati G., Maraschi L., Comastri A., 
  1998, MNRAS, 301, 451
\bibitem[\protect\citeauthoryear{Hardcastle, Birkinshaw \& Worrall}{2001}]{har01}
  Hardcastle M. J., Birkinshaw M., Worrall D. M., 2001, MNRAS, 326, 1499
\bibitem[\protect\citeauthoryear{Hardcastle et al.}{2002}]{har02}
  Hardcastle M. J., Worrall D. M., Birkinshaw M., Laing R. A., 
  Brindle A. H., 2002, MNRAS, 334, 184
\bibitem[\protect\citeauthoryear{Hardcastle et al.}{2003}]{har03}
  Hardcastle M. J., Worrall D. M., Kraft R. P., Forman W. R., 
  Jones C., Murray S. S., 2003, ApJ, 593, 169
\bibitem[\protect\citeauthoryear{Harris \& Krawczynski}{2002}]{harr02}
  Harris D. E., Krawczynski H., 2002, ApJ, 565, 244
\bibitem[\protect\citeauthoryear{Harris \& Krawczynski}{2006}]{harr06}
  Harris D. E., Krawczynski H., 2006, ARA\&A, 44, 463
\bibitem[\protect\citeauthoryear{Inoue \& Takahara}{1996}]{ino96}
  Inoue S., Takahara F., 1996, ApJ, 463, 555
\bibitem[\protect\citeauthoryear{Jones et al.}{1996}]{jon96}
  Jones D. L. et al., 1996, ApJ, 466, L63
\bibitem[\protect\citeauthoryear{Kraft et al.}{2002}]{kra02}
  Kraft R. P., Forman W. R., Jones D., Murray S. S., Hardcastle M. J., 
  Worrall K. M., 2002, ApJ, 569, 54
\bibitem[\protect\citeauthoryear{Kusunose, Takahara \& Li}{2000}]{kus00}
  Kusunose M., Takahara F., Li, H., 2000, ApJ, 536, 299
\bibitem[\protect\citeauthoryear{Marshall et al.}{2002}]{mar02}
  Marshall H. L., Miller B. P., Davis D. S., Perlman E. S.,
  Wise M., Danizares D. R., Harris D. E., 2002, ApJ, 564, 683
\bibitem[\protect\citeauthoryear{Rybicki \& Lightman}{1979}]{ryb79}
  Rybicki, G.B., Lightman, A., 1979, Radiative Processes in Astrophysics, 
  New York, Wiley 
\bibitem[\protect\citeauthoryear{Siemiginowska et al.}{2002}]{sie02}
  Siemiginowska A., Bechtolod J., Aldcroft T. L., Elvis M., Harris D. E.
  Dobrzycki A., 2002, ApJ, 570, 543
\bibitem[\protect\citeauthoryear{Siemiginowska et al.}{2007}]{sie07}
  Siemiginowska A., Stawrz \L., Cheung C. C., Harris D. E., 
  Sikora M., Aldcroft T. L., Bechtold J., 2007, ApJ, 657, 145
\bibitem[\protect\citeauthoryear{Stanghellini et al.}{1998}]{sta98}
  Stanghellini C., O'Dea C. P., Dallacasa D., Baum S. A., Fanti R., 
  Fanti C., 1998, A\&AS, 131, 303
\bibitem[\protect\citeauthoryear{Tingay et al.}{1998}]{tin98}
  Tingay S. J. et al., 1998, ApJ, 115, 960
\bibitem[\protect\citeauthoryear{Wilson \& Yang}{2002}]{wil02}
  Wilson A. S., Yang Y., 2002, ApJ, 568, 133
\end{thebibliography}
\end{document}